\documentstyle[aps,prl,twocolumn]{revtex}
\newcommand{\taup}{\tau^{\prime}}
\newcommand{\alphap}{\alpha^{\prime}}

\begin{document}           % End of preamble and beginning of text.

\title{\bf
Mesoscopic Transport Through Ballistic Cavities: \\
A Random S-Matrix Theory Approach
}

\author{Harold U. Baranger$^{1}$ and Pier A. Mello$^{2}$}

\address{$^{1}$ AT\&T Bell Laboratories 1D-230, 600 Mountain Avenue,
Murray Hill, New Jersey 07974-0636 }

\address{$^{2}$ Instituto de Fisica, Universidad Nacional
Aut\'{o}noma de M\'{e}xico, 01000 M\'{e}xico D.F., M\'{e}xico}

\date{Submitted to Phys. Rev. Lett., March 15, 1994}

\maketitle

\begin{abstract}
We deduce the effects of quantum interference on the conductance of chaotic
cavities by using a statistical ansatz for the $S$ matrix. Assuming that
the circular ensembles describe the $S$ matrix of a chaotic cavity, we find
that the conductance fluctuation and weak-localization magnitudes are
universal: they are independent of the size and shape of the cavity if the
number of incoming modes, $N$, is large.  The limit of small $N$ is more
relevant experimentally; here we calculate the full distribution of the
conductance and find striking differences as $N$ changes or a magnetic
field is applied.
\end{abstract}

\medskip
\narrowtext

The effect of quantum interference on transport through microstructures has
been intensively investigated and is one of the main
subjects of mesoscopic physics \cite{RevMes}. For diffusive transport in
disordered
structures, both microscopic perturbative and macroscopic random matrix
theories give a good account of the phenomena.
In the latter case \cite{RevRMTmes}, one assumes that the transfer
matrix for an
ensemble of such disordered microstructures can be chosen from a simple
statistical ensemble with only symmetry constraints applied. The success
of this theory is perhaps the best theoretical demonstration of the
ubiquity of mesoscopic interference effects.

More recently, interest has focused on quantum transport in ballistic
microstructures--- structures in which impurity scattering can be neglected
so that only scattering from the boundaries of the conducting region is
important \cite{RevMes}. Quantum interference effects
in such structures depend on the nature of the classical
dynamics \cite{Jal90,Oak92,Jen91,Dor91,Lin93,Blu88},
in particular whether it is regular or chaotic \cite{RevCha}.
Recent experiments have studied transport in such ballistic structures
\cite{Mar92,MJBer93,Kel93,Weiss93} and have
detected a difference between nominally regular and chaotic
structures \cite{Mar92}.

The theoretical work on this subject
\cite{Jal90,Oak92,Jen91,Dor91,Lin93} has concentrated on
either numerical quantum calculations or semiclassical theory.
On the other hand, it has been proposed \cite{Blu88} that chaotic
scattering in the quantum regime \cite{QMSca}
should be described by a random matrix theory for the $S$-matrix.
The emphasis in both that work and very recent work on the $S$-matrix
of disordered structures \cite{Jalunp} is on the properties of the
eigenphases of $S$. The eigenphases, however, are not directly connected
to the transport properties because they involve both reflection and
transmission. In contrast, in this
paper we derive the implications of such a random $S$-matrix theory for the
quantum transport properties and provide numerical evidence that this
theory applies to the class of ballistic microstructures
investigated experimentally. In this way we obtain experimentally
accessible predictions for the quantum transport properties of chaotic
billiards.

A quantum scattering problem is described by its $S$-matrix. For scattering
involving two leads (see Fig. \ref{fig:varwlN}) each with $N$ channels
and width $W$, we have
\begin{equation}
S = \left( \begin{array}{cc}
r & t^{\prime} \\
t & r^{\prime}
\end{array} \right)
\label{defS}
\end{equation}
where $r,t$ are the $N \times N$ reflection and transmission matrices for
particles
from the left and $r^{\prime},t^{\prime}$ for those from the right. In terms
of $S$, the conductance is \cite{LanBut}
\begin{equation}
G = (e^2 /h) T = (e^2 /h) Tr \{ tt^{\dag} \} .
\label{defG}
\end{equation}
Because of current conservation $S$ is unitary, $SS^{\dag} = I$, and in
the absence of a magnetic field it is symmetric because of time-reversal
symmetry, $S=S^T$.

We concentrate on situations where the
statistics of the scattering
can be described by an ensemble of $S$-matrices that
assign to $S$ an ``equal a priori distribution'' once the symmetry
restrictions have been imposed. In particular, the possibility of
``direct'' processes caused, for instance, by short trajectories and giving
rise to a nonvanishing averaged $S$-matrix
\cite{Mel85,Blu88}, is ruled out.
The appropriate ensembles are
well-known in classical random matrix theory \cite{RevRMT}
and are called the Circular Orthogonal Ensemble (COE, $\beta=1$) in the
presence of time-reversal symmetry and the Circular Unitary Ensemble (CUE,
$\beta=2$) in the absence of such symmetry. These ensembles are defined
through their invariant measure: the measure on the matrix space which is
invariant under the appropriate symmetry operations.
To be precise \cite{RevRMT}, $d\mu (S) = d\mu (S^{\prime})$,
%\begin{equation}
%d\mu (S) = d\mu (S^{\prime}),
%\label{defmu}
%\end{equation}
where $S^{\prime} = U_0 S V_0$, and $U_0$, $V_0$ are arbitrary fixed unitary
matrices in the case of the CUE with the restriction
$V_0 =U_0^T$ in the COE.
Numerical evidence for the validity of this random matrix theory for
describing quantum-chaotic scattering can be found in Ref. \cite{Blu88}.

Perhaps the most widely studied mesoscopic transport effects are the
magnitude of the conductance fluctuations--- how much the conductance
varies as a magnetic field or gate voltage is applied--- and the size of the
weak-localization correction to the average conductance at $B=0$ \cite{RevMes}.
We therefore start by deriving $\langle T \rangle$ and
$var(T)$ where we use an integration
over the invariant measure
%Eq. (\ref{defmu})
as the average. Such
integrals have been evaluated previously \cite{Mel8090},
and we find that
%\begin{mathletters}
%\label{avgS}
%\begin{equation}
\[
\int d\mu (S) | t_{ab} |^2 =
\frac{ 1 }{ 2N + \delta_{1\beta} }
\]
%\end{equation}
%\begin{equation}
\[
\int d\mu (S) | t_{ab} |^2 | t_{cd} |^2 =
\frac{ 2(N+\delta_{1\beta})(1+\delta_{ac}\delta_{bd}) -
\delta_{ac} - \delta_{bd} }
{ 2N(2N+1)(2N-1+4\delta_{1\beta} ) } .
\]
%\end{equation}
%\end{mathletters}
Performing the trace over channels in Eq. (\ref{defG}), we obtain
\begin{mathletters}
\label{varwlN}
\begin{equation}
\langle T \rangle - N/2 = -\delta_{1\beta} N/(4N+2)
\rightarrow (-1/4) \delta_{1\beta}
\label{wlN}
\end{equation}
\begin{equation}
var(T) =
\left\{
\begin{array}{lll}
\parbox{1.2in}{ \vspace{-\abovedisplayskip} \[
\frac{ N(N+1)^2 }{ (2N+1)^2(2N+3) }
\] \vspace{-\belowdisplayskip} }
& \parbox{0.4in}{ \vspace{-\abovedisplayskip} \[
\rightarrow \frac{1}{8} \; ,
\] \vspace{-\belowdisplayskip} }
& \mbox{COE} \\
\parbox{0.7in}{ \vspace{-\abovedisplayskip} \[
\frac{ N^2 }{ 4(4N^2-1) }
\] \vspace{-\belowdisplayskip} }
& \parbox{0.4in}{ \vspace{-\abovedisplayskip} \[
\rightarrow \frac{1}{16} \; ,
\] \vspace{-\belowdisplayskip} }
& \mbox{CUE}
\end{array}
\right.
\label{varN}
\end{equation}
\end{mathletters}
where the limit is as $N \rightarrow \infty$.

We make several comments concerning these results.
(1) Previously, semiclassical theory and numerical calculations
suggested that the weak-localization correction, $\langle T \rangle - N/2$,
and the magnitude of the conductance fluctuations, $var(T)$, are independent
of the size of the system for chaotic billiards \cite{Jal90}.
This is the analogue of the ``universality'' of the conductance
fluctuations in the diffusive regime \cite{RevMes}.
{\it Since the number of modes is proportional to the size of the system
($N = int[kW/\pi]$), our $N \rightarrow \infty$ results show that the
conductance fluctuations and weak-localization are universal within
the random $S$-matrix theory.}
(2) In the large $N$ limit, $var(T)$ in the presence of time-reversal symmetry
is twice as large as in the absence of symmetry, as for diffusive
conductance fluctuations. This demonstrates the universal effect of symmetry.
(3) Both quantities show some variation in the small $N$ regime typical of the
experiments \cite{Mar92,MJBer93,Kel93}.
Note, for instance, that the ratio of $var(T)$ in the presence
and absence of symmetry is not $2$ for small $N$.
(4) The values obtained in the $N \rightarrow \infty$ limit
are the same as those obtained from a random matrix theory for the
Hamiltonian \cite{Iid90} in which one assumes that the Hamiltonian of the
billiard is
described by the Gaussian Ensembles and finds the conductance by
coupling the billiard to leads in a random way. Quantitative agreement
between
$S$-matrix and Hamiltonian random matrix theories has been noted in the
past \cite{Mel85} but is not fully understood.

The predictions of the random matrix theory are compared to the conductance
of a stadium billiard in Fig. \ref{fig:varwlN} computed using the method
of Ref. \cite{Bar91}.
In these calculations, the $S$-matrix varies as a function of
energy because of the
resonances occurring in the cavity. We estimate that
the resonances are moderately overlapping for $N=1$ and that the
width to spacing ratio increases linearly with $N$. The
basic assumption ({\it ergodic hypothesis}) is that through these
fluctuations $S$ covers the
matrix space with uniform probability. This should apply to
billiards in which the effect of short non-chaotic paths is minimized.
We therefore use a stadium billiard in which (1) a stopper blocks any direct
transmission between the leads, (2) a stopper blocks the whispering
gallery trajectories which hug the half-circle part of the stadium, and
(3) the stadium is asymmetrized to break all reflection symmetries.
We obtain excellent agreement between the energy
averages found numerically and
the invariant-measure ensemble averages introduced above.
In particular, both the variation
at small $N$ and the ratio of $var(T)$ in the presence and absence of
time-reversal symmetry are verified in the billiards.

Motivated by this good agreement, we consider
more detailed predictions of the random $S$-matrix theory: we derive
the full distribution of $T$ for small $N$ and the statistics of the
eigenvalues of $tt^{\dag}$, denoted $\{\tau\}$.
We obtain these results by expressing the invariant measure
%of Eq. (\ref{defmu})
in terms of a set of variables
that includes the $\{\tau\}$. Any unitary matrix of the form in Eq.
(\ref{defS}) can be written as \cite{Mel91}
\begin{equation}
S = \left[ \begin{array}{cc}
v^{(1)} & 0 \\
0 & v^{(2)}
\end{array} \right] \left[ \begin{array}{cc}
- \sqrt{1-\tau} & \sqrt{\tau} \\
\sqrt{\tau} & \sqrt{1-\tau}
\end{array} \right] \left[ \begin{array}{cc}
v^{(3)} & 0 \\
0 & v^{(4)}
\end{array} \right]
\label{polarS}
\end{equation}
where $\tau$ stands for a $N \times N$
diagonal matrix of the eigenvalues $\{\tau\}$
and the $v^{(i)}$ are arbitrary unitary matrices except that
$v^{(3)} = (v^{(1)})^T$ and $v^{(4)} = (v^{(2)})^T$ in the presence of
time-reversal symmetry. Now it is a general property of measures on
vector spaces \cite{Lass50} that a differential arc-length written in the
form
%\begin{equation}
$ d {\sigma}^2 =$ $ \sum_{ab} g_{ab} dx^a dx^b$
%\label{arclen}
%\end{equation}
implies that the volume measure is
%\begin{equation}
$d \mu (V) =$ $\sqrt{det(g)} \prod_a dx^a$ .
%\label{volume}
%\end{equation}
In our case the differential arc-length is simply
$d {\sigma}^2 = Tr \{ dS^{\dag} dS \}$.
%\begin{equation}
%d {\sigma}^2 = Tr \{ dS^{\dag} dS \} .
%\label{defarclenS}
%\end{equation}
Substituting for $S$ the form in Eq. (\ref{polarS}), one finds
($\beta = 1,2$)
\begin{equation}
d\mu (S) = P_{\beta}( \{\tau\} ) \prod_a d\tau_a \prod_i d\mu (v^{(i)})
\label{polarmu}
\end{equation}
where the joint probability distribution of the $\{\tau\}$ is
\begin{mathletters}
\label{jprobtau}
\begin{equation}
P_{2}(\{\tau\}) = C_2 \prod_{a<b} | \tau_a - \tau_b |^2
\end{equation}
\begin{equation}
P_{1}(\{\tau\}) = C_1 \prod_{a<b} | \tau_a - \tau_b |
\prod_c 1/\sqrt{\tau_c} \; ,
\end{equation}
\end{mathletters}
$d\mu (v^{(i)})$ denotes the invariant, or Haar's, measure on the
unitary group \cite{Hamer}, and
$C_{\beta}$ are $N$ dependent normalization constants \cite{Eq10comment}.

The distribution of $T = \sum_{a=1}^{N} \tau_a$ follows from
Eq. (\ref{polarmu}) by integration over the joint probability distribution.
This can be carried out for small $N$; for instance, in the trivial
case $N=1$, $w(T) = 1$ for the CUE and $w(T) = 1/(2\sqrt{T})$ for the COE.
For $N=1-3$ the $w(T)$ derived from the random matrix theory are
plotted in Fig. \ref{fig:distT} and compared to numerical data for billiards.
{\it Note the dramatic
difference between the CUE and COE $w(T)$ in the single mode case, and
the difference within each ensemble between the $N=1$ and $N=2$
cases.} The results for $N=3$ are close to
a Gaussian distribution defined by the two moments given in Eqs.
(\ref{varwlN}).

The agreement between the numerical and theoretical results in
Fig. \ref{fig:distT} is very good in terms of both the difference
between COE and CUE and the dependence on $N$ \cite{COE3}.
These effects should be observable in experiments, which are typically
done in the few mode limit, and would
provide a clear test of the applicability of random $S$-matrix theory
to experimental microstructures.

Though not experimentally accessible, the $\{\tau\}$ are
theoretically interesting because of their fundamental relation to
the conductance.  We obtain further information by writing the
joint probability density in the form
\begin{equation}
P_{\beta}(\{\tau\}) = C_{\beta}
\exp \{
 -\beta [ \sum_{a<b} \ln |\tau_a - \tau_b | + \sum_c V_{\beta}(\tau_c) ]
\}
%e^{ -\beta [ \sum_{a<b} \ln |\tau_a - \tau_b | + \sum_c V_{\beta}(\tau_c) ] }
\label{coulgas}
\end{equation}
with $V_2 (\tau ) = 0$ and $V_1 (\tau )$ = $\case{1}{2} \ln \tau$. This is
exactly the form of the joint density in
the global-maximum-entropy approach to quantum transport in disordered
systems \cite{RevRMTmes} and in the Gaussian ensembles \cite{RevRMT}.
Many statistics of such distributions are known asymptotically as
$N \rightarrow \infty$ \cite{RevRMT,Brez7893,Pol89,Ben93,Ben94}.
For instance, the known form of the asymptotic two-point correlation
function \cite{Brez7893,Ben94} can be used to obtain the asymptotic
value of $var(T)$:
\begin{equation}
var(T) = \int_0^1 d\tau \int_0^1 d\taup \tau \taup \rho_2 (\tau ,\taup )
\rightarrow 1/8\beta .
\label{asympvar}
\end{equation}
This result agrees with Eq. (\ref{varN}) and is independent of the
potential $V (\tau ) $.
Therefore the asymptotic value of $var(T)$ is the same for
a large class of random matrix theories, another aspect of
the ``universality'' of conductance fluctuations \cite{Ben93}.

In the CUE case, the joint density in Eq. (\ref{coulgas}) is of
a special form suitable for the random matrix theory
method of orthogonal polynomials \cite{RevRMT,RevRMTmes}.
In this case, $V=0$ and the eigenvalues are restricted to
$[0,1]$ so the Legendre polynomials are appropriate \cite{Leff64}.
In terms of these polynomials, the exact eigenvalue density and two-point
correlation function are \cite{Leff64,Brez7893}
%\begin{mathletters}
%\label{orthogpoly}
%\begin{equation}
\[
\rho (\tau )  = \frac{N^2}
{4 \tau (1-\tau )}
[ P_N^2 (\alpha )-2\alpha P_N (\alpha )P_{N-1}(\alpha )+P_{N-1}^2 (\alpha )]
\]
%\end{equation}
\begin{eqnarray}
%\begin{eqnarray*}
\lefteqn{ \rho_2 (\tau ,\taup )  = \rho(\tau ) \delta (\tau - \taup )/N } \\
 & & - [P_N(\alpha )P_{N-1}(\alphap ) - P_{N-1}(\alpha )P_N(\alphap )]^2 /
16(\tau - \taup )^2 \nonumber
\end{eqnarray}
%\end{mathletters}
where $\alpha \equiv 2\tau - 1$.
Using the asymptotic expansion of the Legendre polynomials as
$N \rightarrow \infty$ and some
smoothing, one finds that
$\rho (\tau ) \rightarrow$ $N/\pi \sqrt{\tau (1-\tau )}$ \cite{Leff64}
and recovers the expression in Refs. \cite{Brez7893} and \cite{Ben94} for
the asymptotic two-point correlation function.
Previous work has shown that the statistics of the eigenvalues $\{\tau\}$
follows that of the Gaussian unitary ensemble in the large $N$
limit \cite{Leff64}.
Comparison of the predictions of the last two paragraphs with
numerical results further supports the applicability of the COE and CUE
to ballistic cavities.

In summary, we have derived the consequences for quantum transport of the
assumption that the $S$-matrix of a chaotic cavity follows the circular
ensembles. We have shown that the magnitude of both the conductance
fluctuations and the weak-localization in a chaotic microstructure are
universal in the large $N$ limit.
The small $N$ limit is most relevant
experimentally, and here we find deviations from the asymptotic behavior
(Fig. \ref{fig:varwlN}) as
well as a striking dependence of the full distribution of $T$ on both
$N$ and magnetic field (Fig. \ref{fig:distT}).
In closing, we emphasize that we have neglected
the ``direct'' scattering due to short paths
($\langle S \rangle = 0$); since such scattering is important in many
chaotic cavities, the effect of these processes on quantum transport
remains an important open question, which, in principle, could be investigated
using the information-theoretic model of Ref. \cite{Mel85}.

We thank R. A. Jalabert for several valuable discussions.
One of us (PAM) appreciates the kind hospitality of the
Wissenschaftkolleg zu Berlin where part of this work was performed.

{\it Note added---} While preparing this work for publication we received
a preprint by C. W. J. Beenakker, R. A. Jalabert, and J.-L. Pichard
which contains some overlapping material. We thank J.-L. Pichard for
sending us this preprint.

\begin{figure}
\caption{
The magnitude of the weak-localization correction [panel (a)]
and the conductance fluctuations [panel (b)]
as a function of the number of modes in the leads, $N$.
The numerical results for $B=0$ (squares with statistical error bars) agree
with the prediction of the
COE (diamonds connected by dotted line), while those for $B \neq 0$ (triangles)
agree with the prediction of the CUE (diamonds connected by dashed line).
The inset shows a typical ballistic cavity used. The numerical
results involve averaging over (1) energy at fixed $N$ (50 points), (2)
6 different cavities obtained by changing the two stoppers,
and (3) 2 magnetic fields for $B \neq 0$ ($BA/\phi_0 = 2$, $4$ where $A$ is
the area of the cavity).
}
\label{fig:varwlN}
\end{figure}

\begin{figure}
\caption{
The distribution of the transmission intensity at fixed $N =$ 1, 2, or 3
both in the absence (first column) and presence (second column) of a
magnetic field. The numerical results (plusses with statistical error bars)
are in good agreement
with the predictions of the circular ensembles (dashed lines). Note in
particular the striking difference between the $N=1$ (first row) and
$N=2$ (second row) results and between the $B=0$ and $B \neq 0 $ results for
$N=1$. For $N=3$ the distribution approaches a Gaussian (dotted lines).
The cavities used are the same as those in Fig. 1; for
$B \neq 0$, $BA/\phi_0 =$ 2, 3, 4, and 5 were used.
}
\label{fig:distT}
\end{figure}


\begin{references}
\bibitem{Eq10comment}
In Eq. (\ref{polarmu}) we have omitted phase factors which reflect the
arbitrariness of the decomposition Eq. (\ref{polarS}) for the CUE case.
None of the results reported here are affected; see Ref. \cite{Mel91} for
a full discussion. Also note that $\prod_i$ runs from $i=1$ to $2$ for
the COE and $i=1$ to $4$ for the CUE.
\bibitem{COE3}
The agreement is poorest in the COE $N=3$ case. Further analysis is necessary
to determine whether there is a systematic deviation from the theory near
$T=1.25$ in this case.
\end{references}
\end{document}